\documentclass{article}
\usepackage{graphicx}  
\usepackage{amsmath}   
\usepackage{amssymb}   
\usepackage[export]{adjustbox}
\usepackage{bm} 
\usepackage{dcolumn}
\usepackage{color}
\usepackage{mathrsfs}
\usepackage{amsfonts}
\usepackage{varioref}
\RequirePackage[colorlinks,citecolor=blue,urlcolor=magenta,linkcolor=blue]{hyperref}

\labelformat{section}{Section #1} 
\labelformat{subsection}{Section #1} 
\labelformat{subsubsection}{Section #1}
\labelformat{subsubsubsection}{Section #1}
\labelformat{equation}{Eq.~(#1)} 
\labelformat{figure}{Fig.~#1} 
\labelformat{subfigure}{Fig.~\thefigure#1} 
\labelformat{table}{Tab.~#1} 
\labelformat{appendix}{Appendix #1}

\title{\bf Cosmological screening and the phantom braneworld model}
\author{\bf Sourav Bhattacharya$^{1}$\footnote{sbhatta@iitrpr.ac.in},~~ Stefanos R Kousvos$^{2}$\footnote{skousvos@physics.uoc.gr} ,~~ Stylianos Romanopoulos$^{2}$\footnote{sromanop@physics.uoc.gr}\\
\bf ~~and~Theodore N Tomaras$^{2}$\footnote{tomaras@physics.uoc.gr}\\
$^{1}$\small{Department of Physics, Indian Institute of Technology Ropar, Rupnagar, Punjab 140 001, India}\\
$^{2}$\small{ITCP and Department of Physics, University of Crete, 700 13 Heraklion, Greece} }

\begin{document}
  
\maketitle
\begin{abstract}
\noindent
The scalar and vector cosmological perturbations at all length scales of our Universe are studied in the
framework of the phantom braneworld model. The model is characterized by the parameter $\Omega_M\equiv M^3/2m^2H_0$, with $M$ and $m$ the 5- and 4-dimensional Planck scales, respectively, and $H_0$ the Hubble parameter today, while $\Omega_M\rightarrow 0$ recovers the $\Lambda\rm CDM$ model. Ignoring the backreaction due to the peculiar velocities and also the bulk cosmological constant, allows the explicit computation of the gravitational
potentials, $\Phi$ and $\Psi$.  They exhibit exponentially decreasing screening behaviour characterized by a screening length which is a function of the quasidensity parameter $ \Omega_M$. 
\end{abstract}
\vskip .5cm
\noindent
\noindent
{\bf keywords :} Braneworld model, scalar-vector perturbations, cosmological screening 
\bigskip
\section{Introduction}\label{s1}
\noindent
In the braneworld (BW) model  the  $3+1-$ dimensional Universe we live in is a timelike hypersurface (the brane) of codimension one or more, embedded in a higher dimensional spacetime (the world), see~\cite{Brax:2004xh, Maartens:2010ar}  for a vast review and also references therein. Unlike the higher dimensional theories such as Gauss-Bonnet gravity, e.g.~\cite{Maeda:2006hj}, in the BW model {\it all} standard model matter fields  are confined on the brane whereas only gravity can propagate in the extra dimension(s).

The existence of the extra dimension implies departure from General Relativity. For example in the  Randall-Sundrum model with a single extra dimension, the modification occurs  at the small scales~\cite{Randall:1999ee, Randall:1999vf}. The extra dimension needs to neither be small nor compact and can even be infinite. Compact extra dimensions, on the other hand, imply an infinite and discrete Kaluza-Klein spectrum on the brane, see e.g.~\cite{Durrer:2003rg}. We further refer our reader to ~\cite{Mak:2004hv}-\cite{Harko:2007yq} for a description of fitting the galaxy rotation curves and the study of gravitational lensing in this model.  While the extra dimension  is usually taken to be spacelike, we refer our reader to~\cite{Shtanov:2002mb} for a timelike extra dimension. 

Discussions on static solutions such as a black hole in the BW model can be seen in ~\cite{Kanti:2013lca, Kanti:2015poa, Nakonieczna:2016qlm, Kar:2015fva} and  references therein.  For the so called two branch RS-I model, from the modification of  Newton's law, the upper bound on the bulk anti-de Sitter radius turns out to be $l\lesssim 14\, \mu m$; whereas for the one branch RS-II model, the binary gravity wave data puts a bound : $l\lesssim 3.9\, \mu m$~\cite{Yagi:2016jml}. Probing the extra dimensional effects by studying the strong gravitational lensing can be seen in~\cite{Chakraborty:2016lxo}. 
We refer our reader to~\cite{Kiritsis:2002ca} for a modification of the RS model with cosmological constants associated with both the bulk and the brane, fine tuned to make the bulk flat. This scenario is in particular helpful to estimate the energy lost by the brane via the Kaluza-Klein gravitons. In~\cite{Kiritsis:2002zf, Kofinas:2005hc}, the effect of brane - bulk energy exchange on cosmology was investigated and a model where our current universe is obtained as a late time attractor was proposed. 
We further refer our reader to~\cite{CliftonEtAl2012} for a vast review and an exhaustive list of references pertaining to gravity and cosmology in the context of the braneworld model.  

In this paper, we shall be interested in an extension of the Dvali-Gabadadze-Porrati braneworld (DGP) model~\cite{Collins:2000yb}-\cite{Shtanov:2000vr} containing in the action, the $4$-dimensional Ricci scalar on the brane, induced by the one loop correction due to the graviton-matter interaction, and the extrinsic curvature of the brane. This model, unlike the Randall-Sundrum case, modifies gravity only beyond a characteristic length scale, depending on the five- and four-dimensional Newton constants. The relevant equation of motion gives rise to two branches of cosmological solutions, both with flat spatial sections, one being self accelerated without requiring any dark energy/cosmological constant, whereas the other branch (the normal branch) requires at least one cosmological constant to accommodate for the current accelerated expansion~\cite{Deffayet:2000uy, Deffayet:2001pu, Charmousis:2006pn}. However, the former was shown to have ghost instability in subsequent works~\cite{Gorbunov:2005zk, Koyama:2007zz}, leaving only the ``normal" branch to be a possible alternative to the $\Lambda{\rm CDM}$ model.  

Furthermore, the equation of state parameter for the effective dark energy source is time dependent, $w(t)$, and turns out to be less than minus one today ~\cite{Sahni:2002dx}-\cite{Shtanov:2005}. For a certain range of parameter values, $w(t)$ will reach asymptotically the value $-1$ (the de Sitter phase). Otherwise, the universe can even leave at some stage the phase of accelerated expansion reentering matter domination, thus evading the so called phantom disaster~\cite{Weinberg:2008zzc}. Since  $w(t)<-1$ in the current epoch, this model is often called ``phantom braneworld model". Interestingly, this model indicates that the expansion of our universe was stopped at redshift $z\gtrsim 6$ and `loitered' there for a long period of time favouring structure formation. Arguments supporting this, based on the observed data of population of the quasistellar objects and supermassive black holes in $6\lesssim z\lesssim 20$ can be found in~\cite{Alam:2005pb}. Scalar cosmological perturbation theory in the phantom braneworld model and further details are studied in~\cite{Viznyuk:2012oda, Bag:2016tvc}, while in ~\cite{Bhattacharya:2017ydy} the stability analysis of large scale cosmic structures via their size-versus-mass study in the context of the present model and in the presence of a bulk cosmological constant, was performed. We also refer our reader to~\cite{Visinelli:2017bny} for constraints on the braneworld model via gravity wave data. See also~\cite{Vagnozzi:2018jhn} for phenomenological arguments in the favour of $w(t)<-1 $ via the neutrino mass higherarchy. 

The braneworld model under study here is assumed to have `zero thickness' in the extra dimension. Interesting effects however, may arise when one considers a thick brane~\cite{Kiritsis:2001bc, Kiritsis:2001mkv}. In particular, in such a scenario, with a large extra dimension, one can have a new energy scale on the brane, determined by both brane thickness and the size of the extra dimension. For energies much larger than this new scale, the physics in the brane depends upon the position along the extra dimension, while for much smaller energies the equivalence principle may be violated, resulting in certain fine tuning to preserve it. 

Given that the phantom braneworld model modifies gravity significantly at large scales, it becomes an interesting task to investigate this model's prediction at arbitrarily large distances. One such arena seems to be the study of screening effects, where certain terms in the scalar perturbation equation, which we can ignore at small scales, lead to modifications of the gravitational potential at large scales~\cite{Eingorn:2015hza}-\cite{Wang:2017krj}. By approximating the inhomogeneities  of our universe as delta function sources, a first order analytical formalism  for the cosmological scalar and vector perturbations for the ${\Lambda{\rm CDM}}$ model was developed recently in~\cite{Eingorn:2015hza}, where a Yukawa-like fall-off of the gravitational potential was derived at large scales. Various extensions of this work, including the case of interacting fluid sources, can be found in~\cite{Eingorn:2015yra, Burgazli:2015mzm, Zhuk:2016imt, Bouhmadi-Lopez:2016cja, Eingorn:2016kdt, Eingorn:2017adg}. Discussions on the  $N$-body simulations in the context of cosmic screening can be seen in~\cite{Hahn:2016roq, Fidler:2017pnb}.
We further refer our reader to~\cite{Brilenkov:2017gro, Wang:2017krj} for second order computations on the scalar perturbation pertaining respectively to the ${\Lambda{\rm CDM}}$ and the Einstein de Sitter models. The extra dimensional scenario is certainly not included in the above examples. Motivated by this, we shall study in this work the first order cosmological screening in the phantom braneworld model.  Our chief goal would be, apart from casting the perturbation equations in a suitable form and solving them, to point out differences of this model from $\Lambda{\rm CDM}$, that can arise at very large scales.

The paper is organized as follows. In the next section we briefly review the phantom braneworld model. In \ref{s3} we develop the first order equations pertaining to the scalar and the vector perturbations with no bulk cosmological constant. In \ref{s4} we solve for the scalar perturbation ignoring the peculiar velocities, and compare it both analytically and numerically with the ${\Lambda {\rm CDM}}$ model. We conclude with a discussion \ref{s5}.   

We shall use mostly negative signature for the metric and will set $c=1$ throughout. 

\section{The phantom braneworld model}\label{s2}
\noindent
Let us first briefly review the basic features of the phantom braneworld model, details of which can be seen in e.g.~\cite{Bag:2016tvc} and references therein. The relevant action  is given by,
\begin{equation}\label{eq1}
S=M^3\Bigg[\int_{\rm bulk}\Big(\mathcal{R}-2\Lambda_{5D}\Big)-2\int_{\rm brane}K\Bigg]+\int_{\rm brane}\Big(m^2 R-2\sigma\Big)-\int_{\rm brane}L(g_{\mu\nu},\phi)
\end{equation}
where $\mathcal{R}$ and $R$ are the Ricci scalars corresponding to five (the bulk) and four dimensions (the brane) and $M$ and $m$ are the respective Planck masses. The quantity $\Lambda_{5D}$ is the cosmological constant in the bulk and $\sigma$ is the brane tension, related to the  brane cosmological constant $\Lambda$ by $\Lambda=\sigma/m^2$. $K$ is the trace of the extrinsic curvature of the brane.  $L(g_{\mu\nu},\phi)$ stands collectively for all matter fields, $\phi$, confined to the brane and $g_{\mu \nu}$ is the induced metric on it. For our current purpose, $\phi$ would correspond only to the cold dark matter.

Being interested in the 3+1-dimensional physics, we choose to measure energies in units of the 4-dimensional Planck mass $m$. So, we set $m=1$ throughout.

Using the  Gauss-Codacci relations, the  Einstein equations on the brane become
\begin{equation}\label{eq2}
G_{\mu\nu}-\bigg(\dfrac{\Lambda_{RS}}{b+1}\bigg)g_{\mu\nu}=\bigg(\dfrac{b}{b+1}\bigg) T_{\mu\nu}-\bigg(\dfrac{1}{b+1}\bigg)\bigg[\dfrac{1}{M^6}Q_{\mu\nu}-\mathcal{C}_{\mu\nu}\bigg]
\end{equation}
where
\begin{equation}\label{eq3}
b=\frac{1}{6} \Lambda l^2 \,,\quad l=\dfrac{2}{M^3}\,,\quad \Lambda_{RS}= \frac{\Lambda_{5D}}{2}+\frac{1}{12} \Lambda^2 l^2
\end{equation}
are convenient parameters, and
\begin{equation}\label{eq4}
Q_{\mu\nu}=\dfrac{1}{3}EE_{\mu\nu}-E_{\mu\lambda}E^\lambda_{\ \nu}+\dfrac{1}{2}\Big(E_{\rho\lambda}E^{\rho\lambda}-\dfrac{1}{3}E^2\Big)g_{\mu\nu}, \qquad
	E_{\mu\nu}\equiv G_{\mu\nu}-T_{\mu\nu}, \qquad E=E^\mu_{\ \mu}
\end{equation}
The  tensor $\mathcal{C}_{\mu\nu}$ is traceless, coming from the projection of the five-dimensional Weyl tensor onto the brane. Taking the divergence of \ref{eq2} yields the constraint equation,
\begin{equation}\label{eq5}
\nabla^{\mu}\Big(\frac{b+1}{m^2} E_{\mu\nu}+ T_{\mu\nu}+\frac{1}{M^6} Q_{\mu\nu}-  \mathcal{C}_{\mu\nu} \Big)=0
\end{equation}
 The spatially homogeneous Einstein equation reads (in conformal time, $\eta$) with the cold dark matter as the source, 
\begin{equation}\label{eq6}
\dfrac{\mathcal{H}^2}{a^2}=\dfrac{\bar{\rho}}{3 a^3}+\dfrac{\Lambda}{3}+\dfrac{2}{l^2}\Bigg[1-\sqrt{1+l^2\bigg(\dfrac{\bar{\rho}}{3 a^3}+\dfrac{\Lambda}{3}-\dfrac{\Lambda_{5D}}{6}-\dfrac{C}{a^4}\bigg)}\Bigg]
\end{equation}
where $a\equiv a(\eta)$ is the scale factor, ${\cal H}= a^{-1}da/d\eta$ is the Hubble rate and ${\bar \rho}$ the time independent background homogeneous cold dark matter density in co-moving coordinates. The constant $C$ is due to the existence of the Weyl tensor in the bulk.  Due to the radiation like behavior of the term containing $C$, it is often named ``Weyl radiation". We shall ignore its backreaction effects onto the cosmological background, though we shall take into account the inhomogeneous perturbations of the projection of the Weyl tensor. We will also ignore the backreaction effects of $\Lambda_{5D}$. Taking $\Omega_M \to 0$ in the above equation one recovers the $\Lambda{\rm CDM}$ limit. Notice that \ref{eq6} in the absence of $ \Lambda_{5D}$ and $ C$ may be conveniently expressed as 
\begin{equation}\label{eq8}
\frac{\mathcal{H}}{a}=\frac{1}{l}\left[ \sqrt{1+ \frac{l^2}{3}\left( \frac{\bar{\rho}}{a^3} +\Lambda\right)}-1 \right]
\end{equation}

We will also need the derivative of this equation with respect to conformal time $ \eta$

\begin{equation}\label{eq9}
\frac{d\mathcal{H}}{d\eta}=\mathcal{H}^2\Bigg(1-\frac{3\Omega_m}{2(1+\Omega_M)}\Bigg)
\end{equation}
where\footnote{We would like to mention that the quantity $ \Omega_M$ is often defined as $ \sqrt{\Omega_l}$ in the related literature} 
\begin{equation}\label{eq10}
\Omega_m = \frac{\bar{\rho}}{3 \mathcal{H}^2a}, \qquad
	\Omega_M=\frac{aM^3}{2\mathcal{H}}, \qquad \Omega_\sigma = \frac{\Lambda}{3 \mathcal{H}^2}
\end{equation}
Consequently, \ref{eq6} takes the simple form
\begin{equation}
\Omega_m+\Omega_\sigma-2\Omega_M=1
\label{omega}
\end{equation}
in everything that follows we have replaced $ \bar{\rho}$ in favor of $ 3\mathcal{H}^2 \Omega_m a$, $ M^3$ in favor of $ 2\Omega_M \mathcal{H}/a$ and $ \Lambda$ in favor of $3\mathcal{H}^2 \Omega_\sigma$. It is very convenient since everything we derive may be expressed as functions of $ \Omega_m$ and $ \Omega_M$, only ($ \Omega_\sigma$ is solved for from \ref{omega}). The advantage of this procedure is twofold, firstly these parameters are dimensionless and we claim rather intuitive to handle, secondly these will make comparison to $ \Lambda \rm{CDM}$ trivial by simply taking $ \Omega_M$ to zero.

\section{Derivation of scalar and vector perturbation equations }\label{s3}

\noindent
We shall extend below the linear perturbation scheme developed for  the ${\Lambda}{\rm CDM}$ model in~\cite{Eingorn:2015hza}  to the phantom braneworld model described in the preceeding section. 
We start with the ansatz for the first order McVittie metric on the brane  in the Cartesian coordinates,
\begin{equation}\label{eq11}
ds^2=a^2(\eta) \big[(1+2\Phi(\eta,\bm{x}))d\eta^2+2B_i(\eta,\bm{x}) d\eta dx^i-(1-2\Psi(\eta,\bm{x}))\delta_{ij}dx^idx^j\big]
\end{equation}
where $\Phi$, $\Psi$ and  $B_i$'s are respectively the scalar and vector perturbations and the bold font is used to indicate a vector, which determines the position in space where the potentials are evaluated at. Note that unlike the $\Lambda{\rm CDM}$, $\Phi \neq \Psi$ here, owing to the  anisotropic stresses originating from  the bulk, e.g.~\cite{Bag:2016tvc}. 	

We shall consider the backreaction effects due to $N$ self gravitating moving point masses. Following~\cite{Eingorn:2015hza}, we define the proper interval for the $n$-th  mass,
\begin{equation}\label{eq12}
ds_n=a(\eta) \big[(1+2\Phi)+2B_i v_n^i-(1-2\Psi)\delta_{ij}v_n^iv_n^j\big]^{1/2}d\eta 
\end{equation}
The peculiar velocities  appearing above  can be evaluated by subtracting from the observed  velocity of the mass, the velocity due to the  Hubble flow, e.g.~\cite{Weinberg:2008zzc}.

The energy momentum tensor for these point masses is then given by
\begin{equation}\label{eq13}
T^{\mu\nu}=\sum\limits_n \dfrac{m_n }{\sqrt{-g}}\dfrac{dx^\mu_n}{d\eta}\dfrac{dx^\nu_n}{d\eta}\dfrac{d\eta}{d s_n}\delta^3(\bm{x}-\bm{x}_n)
\end{equation}
where $ \bm{x}_n$ is the value of the $x$ coordinate (as defined in the metric \ref{eq11}) where the $ n^{th}$ particle is located at.
Existing data shows that the peculiar velocities are in general rather small or non-relativistic, at most of the order of  $10^6\,{\rm ms^{-1}}$~\cite{Girardi:1992me}. Putting these all in together, we find from \ref{eq13} the energy momentum tensor up to the first order,
\begin{equation}\label{eq14}
T^{\mu\nu}=	\dfrac{1}{a^5}
			\begin{pmatrix}
			\big(1-2\Phi+3\Psi\big)\rho 	& \sum\limits_n \rho_n   v_n\\
			\sum\limits_n \rho_n  v_n       & 0
			\end{pmatrix}
\end{equation}
where each $\rho_n$ corresponds to a delta function point mass located at $ {\bm r}_n$,
\begin{equation}\label{eq15}
\rho_n\equiv m_n\delta^3(\bm{x}-\bm{x}_n)
\end{equation} 
We decompose the total energy density $\rho$ in \ref{eq14} as,		
\begin{equation}\label{eq16}
\rho=\bar{\rho}+\delta\rho(\eta,\bm{x}) \,, \qquad \bar{\rho}= \sum_n m_n/V
\end{equation}
where $\delta \rho(\eta ,\bm{x})$ stands for the contribution of the inhomogeneities. The index $n$ runs over all $N$ particles in the Universe. Note here that $\delta\rho$ is not treated as a perturbation, due to the fact that it is dominant at small scales (see \cite{Chisari:2011iq}).

Since we  must have $|\Psi|,~|\Phi| \ll 1$ in \ref{eq11}, we write from \ref{eq14} at first order,
\begin{equation}\label{eq17}
\delta T_{00}=\dfrac{1}{a}\big(\delta \rho +2\bar{\rho}\Phi+3\bar{\rho}\Psi\big)\,, \qquad 
\delta T_{0i}=\dfrac{1}{a}\big(\bar{\rho} B_i-\sum\limits_{n}\rho_n  v_n^i\big) \,, \qquad 
\delta T_{ij}=0
\end{equation}
the geodesic equation for the n$^{\rm th}$ particle in \ref{eq12} also reads,
\begin{equation}\label{eq18}
\left(a \bm{B}|_{\bm{x}=\bm{x}_n}-a\bm{v}_n\right)'=a\nabla \Phi|_{\bm{x}=\bm{x}_n}
\end{equation}	
where the `prime' denotes differentiation once with respect to the conformal time $\eta$ and the variations $\delta T_{\mu\nu}$, $\delta {\cal C}_{\mu\nu}$ and $\delta Q_{\mu\nu}$ depend on both space and time. Since we wish to build a perturbation scheme valid all the way to superhorizon scales, we cannot assume that the perturbations' spatial variations dominate over the temporal ones, unlike the case of the study of cosmic structures, e.g.~\cite{Weinberg:2008zzc}. 

Finally, we come to the perturbation of the Weyl tensor's projection onto the brane, $\delta {\cal C}_{\mu\nu}$. Its most generic form is given by, e.g.~\cite{Bag:2016tvc}, 
\begin{equation}\label{eq19}
\delta \mathcal{C}_{\mu\nu}=\dfrac{1}{a^2} 	
		\begin{pmatrix} 	\delta \rho_{\mathcal{C}}& 		\partial_iv_{\mathcal{C}}\\
	                		\partial_iv_{\mathcal{C}}& 		\dfrac{\delta \rho_{\mathcal{C}}}{3}\delta_{ij}-\delta \pi_{ij}
		\end{pmatrix}
\end{equation} 
where $\delta\pi_{ij}=(\nabla_i\nabla_j-g_{ij}\Delta/3)\delta\pi_\mathcal{C}$ ($\Delta$ stands for the Euclidean $3$-Laplacian) is trace free and $\delta \rho_{\cal C}$, $ v_{\cal C}$ and $\delta \pi_{\cal C}$ are scalars. In particular, $ v_{\cal C}$ can be regarded as a momentum potential, whose backreaction effects will also be ignored, while considering its time evolution also negligible.

The Einstein equations on the brane~\ref{eq2}, at first order read, after using \ref{eq17}, \ref{eq19},
\begin{equation}\label{eq20}
\Delta\Psi-\dfrac{9\mathcal{H}^2\Omega_m }{2m_{\rm eff}^2}\Psi-3\mathcal{H}\Psi'-3\mathcal{H}^2\Phi=\dfrac{\delta\rho }{2m_{\rm eff}^2a}+\dfrac{\Omega_M}{2 m_{\rm eff}^2a^2}\delta\rho_\mathcal{C}
\end{equation}
which is the $ {00}$ component, and for $i\neq j$ 
\begin{equation}\label{eq21}
\dfrac{\Omega_M}{m_{\rm eff}^2a^2}\delta\pi_{ij}-\bigg(1-\frac{3\Omega_m}{4m_{\rm eff}^4}\bigg)\bigg(\partial_i\partial_j\left(\Phi-\Psi\right)-\dfrac{1}{2}\partial_{(i}\Big(B_{j)}'+2\mathcal{H}B_{j)}\Big)\bigg)=0 
\end{equation}
We also have for the vector perturbation,
\begin{equation}\label{eq22}
\dfrac{1}{4}\Delta B_i-\dfrac{3\mathcal{H}^2\Omega_m}{2m^2_{\rm eff}}B_i+\partial_i\left(\Psi'+\mathcal{H}\Phi\right)=-\dfrac{1}{2m^2_{\rm eff}a}\sum\limits_{n}\rho_n  v_n^i
\end{equation}
where $\Delta$ as earlier is the Laplacian on the Euclidean $3$-space and also the function $m_{\rm eff}\equiv m_{\rm eff}(\eta)$ has been introduced
\begin{equation}\label{eq23}
m_{\rm eff}^2\equiv 1+\Omega_M
\end{equation}
The $\Lambda{\rm CDM}$ limit in the above equations is obtained by letting $\Omega_M\to 0$ in which case we recover the results of~\cite{Eingorn:2015hza}. 

At small length scales relevant to cosmic structures, the spatial derivatives of the potential in \ref{eq20} dominate over its temporal derivatives and the other effective mass-like terms appearing on the left hand side. Accordingly, at such small scales, \ref{eq20} reduces to the Poisson equation, yielding a gravitational potential falling off as $1/r$, along with a modified Newton's constant~\cite{Bhattacharya:2017ydy}. For $\Lambda{\rm CDM}$ in particular, 
we have $\delta\rho_\mathcal{C}=0$, yielding Newton's potential. However, at length scales much larger than those of cosmic structures, the temporal derivative and the effective mass terms can be comparable and, as we will show in \ref{s4}, this leads to a significant modification in the behavior of the solution of \ref{eq20}, as is expected due to the presence of the mass-like term on its left-hand side.

The divergence of \ref{eq22}, in the Poisson gauge $\partial_i B_i=0$, gives
\begin{equation}\label{eq24}
\Delta\Xi=\partial_i\left(\sum\limits_n \rho_n {v_n}^i\right)
\end{equation}
where $\Xi:=-2m^2_{\rm eff}a (\Psi'+\mathcal{H}\Phi)$. 
The solution of \ref{eq24} is 
\begin{equation}\label{eq25}
\Xi= \frac{1}{4\pi} \sum_n m_n \frac{(\bm{x}-\bm{x}_n)\cdot\bm{v}_n}{|\bm{x-x}_n|^3}
\end{equation}
\ref{eq22} has the same form as the corresponding solved in \cite{Eingorn:2015hza}. 

In this work, we are chiefly interested in distinguishing the phantom braneworld model from $\Lambda{\rm CDM}$ with respect to the cosmological screening,  which is certainly impossible unless we go to very large length scales. Note that at such scales, the backreaction effects due to the peculiar velocities, which are essentially non-relativistic, would be negligible, e.g.~\cite{Girardi:1992me}. Thus for our current purpose, we shall from now on ignore the peculiar velocities (and hence the vector perturbation) throughout. 

\section{Solutions ignoring peculiar velocities}\label{s4}

\noindent
From the definition of $\Xi$ and \ref{eq25} in the limit of irrelevant peculiar velocities we have (since $ \Xi =0$)
\begin{align}\label{eq26}
\Psi'=-\mathcal{H}\Phi
\end{align}
With this equation in hand we can simplify \ref{eq20} 
\begin{equation}\label{eq27}
\Delta\Psi-\dfrac{9\mathcal{H}^2\Omega_m }{2m_{\rm eff}^2}\Psi=\dfrac{\delta\rho }{2m_{\rm eff}^2a}+\dfrac{\Omega_M}{2 m_{\rm eff}^2a^2}\delta\rho_\mathcal{C}
\end{equation}
On the other hand we can write \ref{eq21}
\begin{equation}\label{eq28}
\dfrac{\Omega_M}{m_{\rm eff}^2a^2}\delta\pi_{\mathcal{C}}=\bigg(1-\frac{3\Omega_m}{4m_{\rm eff}^4}\bigg)\left(\Phi-\Psi\right)+constant
\end{equation}
and recall that in a marginally closed universe with a {\it vanishing} bulk cosmological constant, one has ~\cite{Viznyuk:2012oda},
\begin{equation}\label{eq29}
\Delta\delta\pi_\mathcal{C}=\dfrac{\delta\rho_\mathcal{C}}{2}
\end{equation}	
Combining \ref{eq28} and \ref{eq29} to eliminate $\delta \pi_\mathcal{C}$ we get
\begin{equation}\label{eq30}
\dfrac{\Omega_M}{2 m_{\rm eff}^2a^2}\delta\rho_\mathcal{C}=\bigg(1-\frac{3\Omega_m}{4m_{\rm eff}^4}\bigg)\Delta\left(\Phi-\Psi\right)
\end{equation}
In order to solve for $\Phi$ and $\Psi$ we want one more equation. This comes from the spatial component of \ref{eq5}, after using \ref{eq18} and \ref{eq27}, we obtain
\begin{equation}\label{eq31}
\dfrac{\Omega_M}{2 m_{\rm eff}^2a^2}\bigg(1-\frac{3\Omega_m}{2 m_{\rm eff}^2}\bigg)\delta\rho_\mathcal{C}=\Delta\Phi-\bigg(1+\frac{3\Omega_m}{2 m_{\rm eff}^4}\bigg)\Delta\Psi+constant
\end{equation}
We can substitute \ref{eq30} into \ref{eq27} and \ref{eq31} to obtain a system of two equations with only two unknowns, the perturbations $\Phi$ and $\Psi$. The constant in \ref{eq28} and \ref{eq31} has to be zero in order for the potential to be vanishing at infinity. The solution of the system is straightforward 
\begin{equation}\label{eq32}
\Delta\Phi=I\;\Delta\Psi
\end{equation}
and 
\begin{equation}\label{eq33}
\Delta\Psi-\frac{9\mathcal{H}^2\Omega_{m,\rm eff}}{2}\Psi=\frac{\delta\rho}{2m_{\rm eff,\Psi}^2a}
\end{equation}
where 
\begin{equation}\label{eq34}
I\equiv 1+\frac{4\Omega_M}{1+2m_{\rm eff}^2-\frac{3\Omega_m}{2m_{\rm eff}^4}},\qquad\Omega_{m,\rm eff}\equiv\frac{\Omega_m}{m_{\rm eff,\Psi}^2},\quad
 m_{\rm eff,\Psi}^2\equiv m_{\rm eff}^2+(1-I)\Big(m_{\rm eff}^2-\frac{3\Omega_m}{4m_{\rm eff}^2}\Big)
\end{equation}
\ref{eq33} is identical to the one obtained for ${\Lambda{\rm CDM}}$ derived in \cite{Eingorn:2015hza} if we drop the $ _{\rm eff}$ subscripts, it is trivial to solve our equation by comparing with \cite{Eingorn:2015hza}, and using $ _{\rm eff}$ subscripts wherever appropriate. The solution is 
\begin{equation}\label{eq35}
\Psi\big\vert_{\rm many~particle}=\frac{1}{3}-\dfrac{1}{8\pi m_{\rm eff,\Psi}^2 a}\sum_n\dfrac{m_n }{\vert\bm{x}-\bm{x}_n\vert}\; e^{-\frac{\vert\bm{x}-\bm{x}_n\vert}{\lambda}}
\end{equation}
where
\begin{equation}\label{eq36}
\lambda=\sqrt{\frac{2}{9\mathcal{H}^2\Omega_{m,\rm eff}}}
\end{equation}

For a single particle -- a single central over-density -- the solution for the potential $\Psi$, valid for all length scales is 
\begin{equation}\label{eq37}
\Psi\big\vert_{\rm one~particle}=-\dfrac{1}{8\pi m_{\rm eff,\Psi}^2a}\dfrac{m_0 }{r}\; e^{-r/\lambda}
\end{equation}
where $m_0$ is the mass of the central overdensity and $r=|{\bf x}|$.  We have dropped the $\frac{1}{3}$ which is generated by the existence of an infinite number of point particles. We will prove that this occurs naturally when considering an isolated sub-region of the Universe (e.g. the observable Universe) at the end of this section.  The exponential appearing above clearly indicates the suppression of Newton's potential at large scales, originating from the term present in the perturbation equation behaving as an effective mass. Thus the length scale, $\lambda$,  should be interpreted as a {\it screening length}. 

In every case we can find $\Phi$ solving \ref{eq32}
\begin{equation}\label{eq38}
\Phi=I\;\Psi
\end{equation}
		
\begin{figure}[htbp]
\centering
\includegraphics[scale=0.99]{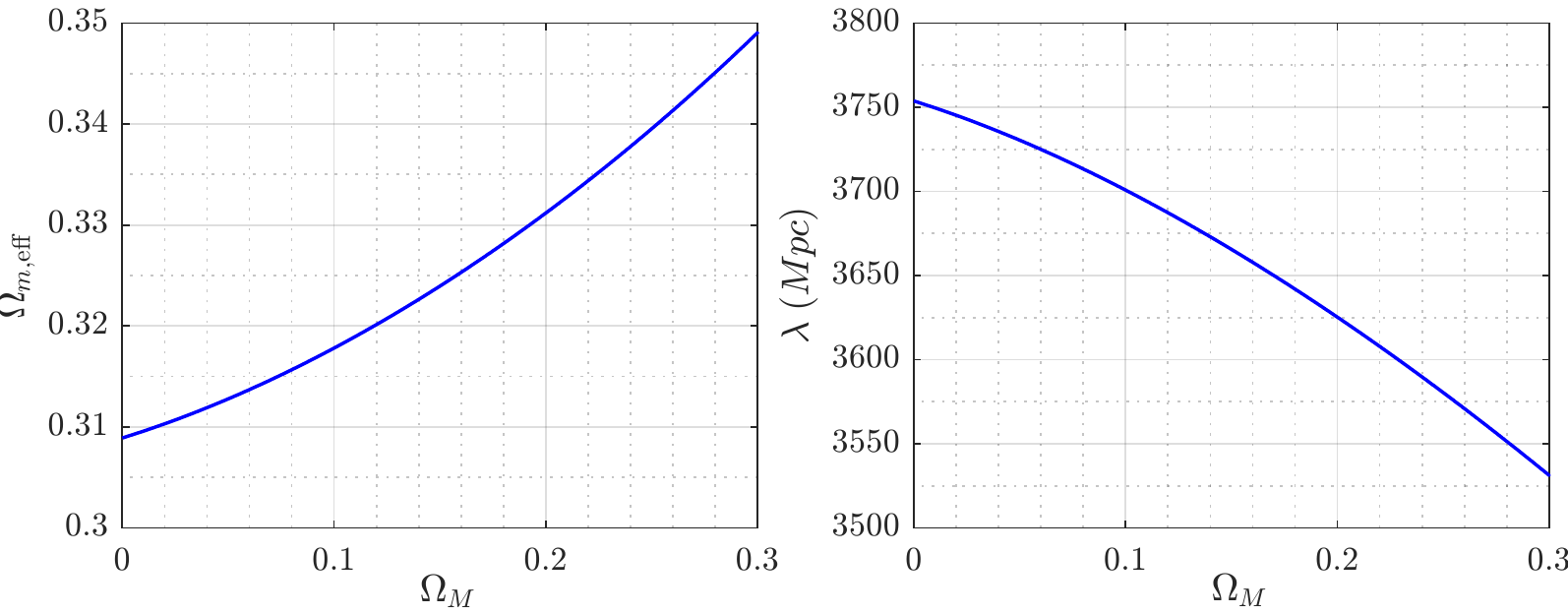}
\caption{Plots of the effective mass density and the screening length versus $\Omega_M$.}
\label{fig1}
\end{figure}

\begin{figure}[htbp]
\centering
\includegraphics[scale=0.99]{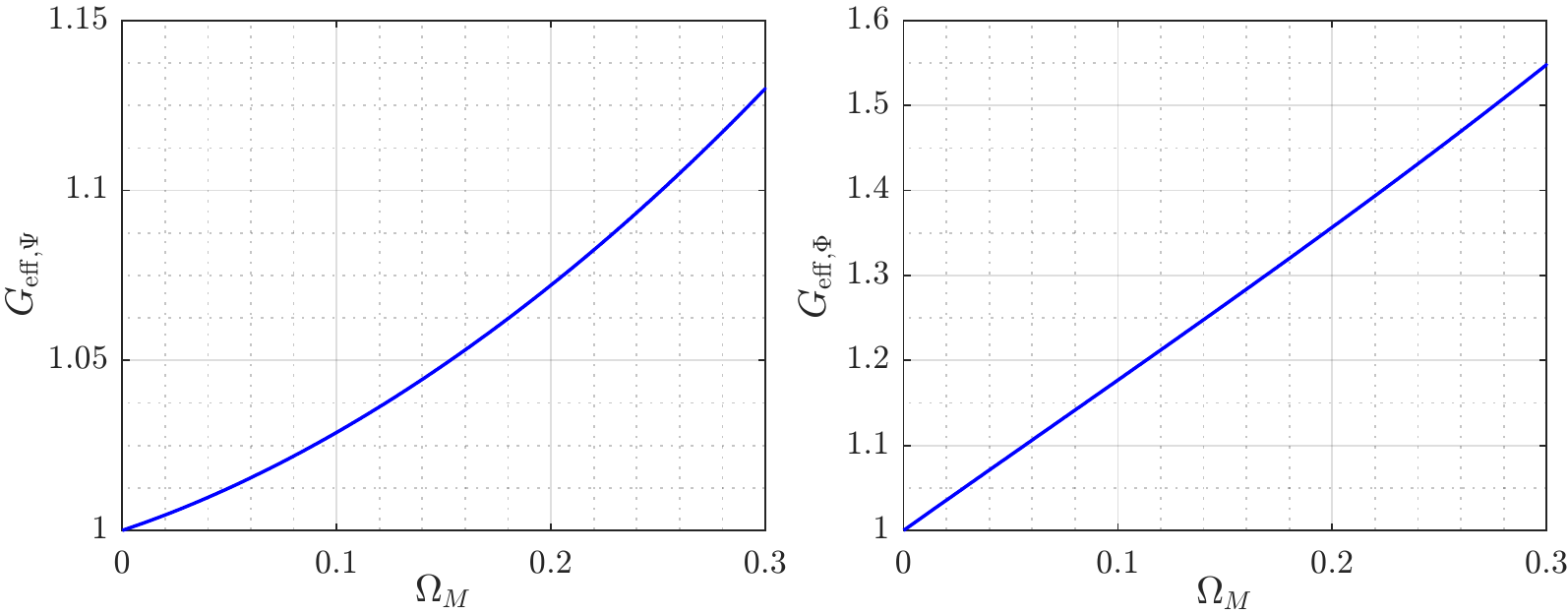}
\caption{ Plots of $1/8\pi m_{\rm eff,\Psi}^2$ and $1/8\pi m_{\rm eff,\Phi}^2$ respectively versus $\Omega_M$. These are proportional to the Newton's constant for each potential respectively. Note that the one for $\Phi$ decreases faster than $\Psi$.} 
\label{fig2}	
\end{figure}

\begin{figure}[htbp]
\centering
\includegraphics[scale=0.99]{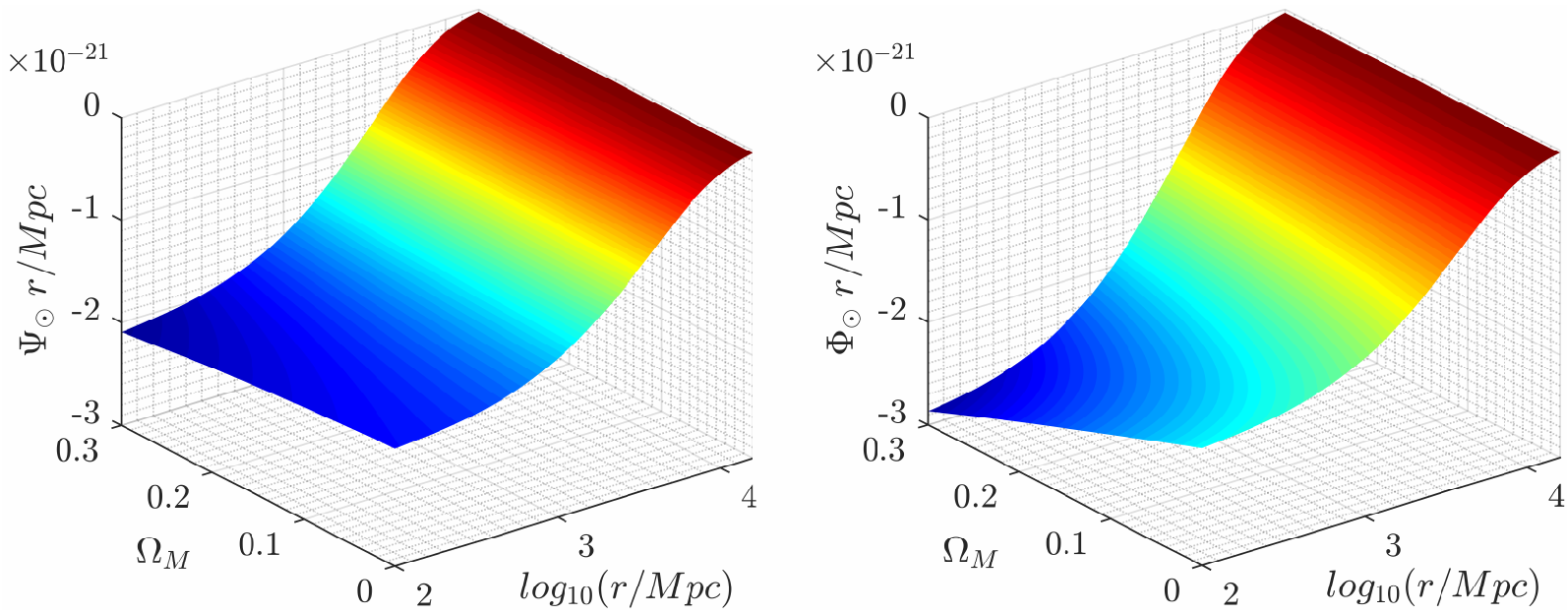}
\caption{  Plot of $\Psi_\odot(r)\, r/Mpc$ and $\Phi_\odot(r)\, r/Mpc$ versus $ \Omega_M$ and $log(r/Mpc)$ for one particle with mass equal to one solar mass.}	
\label{fig3}
\end{figure}

\ref{fig1} $-$ \ref{fig3} elucidate various properties of the gravitational potentials and the screening length. We use the values for cosmological parameters $ \Omega_m=0.3089$ and $H_0=67.74\;km\,s^{-1} Mpc^{-1}$ as specified in \cite{Ade:2013zuv} and we examine the gravitational potentials for one particle with mass $M_\odot=1.989\cdot 10^{30}\;kg$.  \ref{fig1} depicts the behavior of the effective mass density parameter and the screening length versus $ \Omega_M$. The $\Lambda{\rm CDM}$ limit is obtained by letting $\Omega_M\to 0$. 

We also note that since the screening length is typically of the order of ${\cal O}(10^{3})\,{\rm Mpc}$ (\ref{fig1}), at length scales comparable of the size of a typical cosmic structure i.e. ${\cal O}(100)$ Mpc, \ref{eq37} recovers the $1/r$ fall-off of the gravitational potentials. However, the $1/m_{\rm eff,\Phi}^2\equiv I/m_{\rm eff,\Psi}^2$ and $1/m_{\rm eff,\Psi}^2$ terms present modify Newton's `constant' in $\Phi$ and $\Psi$ respectively and make it time dependent, as discussed in~\cite{Bhattacharya:2017ydy}.

\ref{fig2} depicts the behavior of the effective Newton's constant for $\Psi$ and $\Phi$. In the $\Omega_M\to 0$ limit both of them aproach 1 recovering the $\Lambda{\rm CDM}$ limit. Note also that in this limit setting further  $\bar{\rho}\to 0$ ($ \Omega_m \to 0$) removes the exponential fall off since then $\lambda \to \infty $ (cf., \ref{eq36},\ref{eq34}), yielding Newton's potential for a point mass located in a de Sitter universe. It is easy to verify that, as expected, this is the linearized approximation of the Schwarzschild-de Sitter metric in the McVittie coordinate frame. Similar conclusions hold for the potential $\Phi \big\vert_{\rm one~particle}$. Finally, we depict the potentials in \ref{fig3}.


We would now like to show that with respect to the universe {\it visible} to an observer located at some point $\bm{x}$, we can actually get rid of the first term in \ref{eq35}. Indeed, let $N$ be the total number of point sources in \ref{eq35} and let $\widetilde N$ be the number located within the Hubble horizon radius of an observer located at $\bm{x}$. Clearly, we may expect that only these $\widetilde N$ particles would contribute significantly into \ref{eq35}. On the other hand, since we should have $N \to \infty$ in order to obtain a non-vanishing $\bar{\rho}$, it is natural to consider $\widetilde N \ll N$.  
We next split the summations in \ref{eq35} into two parts
$$\sum_{n=1}^{N\to \infty} =\sum_{n=1}^{\widetilde N}+ \sum_{n=\widetilde N+1}^\infty$$
Since the second summation gets contributions from all particles outside the Hubble horizon of the observer, we can average the potential of this part following~\cite{Eingorn:2015hza} and using $\bar{\rho}=\sum_1^{\infty} m_n/V \approx \sum_{\widetilde N+1}^{\infty} m_n/V$.  It is easy to see that this average cancels-out the constant ($1/3$) term in \ref{eq35}, leading to
$$
\Psi\big\vert_{\rm many~particle;\, average}=-\dfrac{1}{8\pi m_{\rm eff,\Psi}^2a}\sum_{n=1}^{\widetilde N}\dfrac{m_n }{\vert\bm{x}-\bm{x}_n\vert}\; e^{-\frac{\vert\bm{x}-\bm{x}_n\vert}{\lambda}}
$$
where ``average'' in the subscript refers to the aforementioned averaging over sources located outside the observer's Hubble horizon. Note that the above formula has a smooth one particle ($\widetilde N=1$) limit, recovering \ref{eq37}.


\section{Discussion}\label{s5}

\noindent
At very large length scales of our universe not decoupled from the cosmic expansion, one might expect  the gravitational  potential to be modified from that of Newton's, as has explicitly been demonstrated for the $\Lambda{\rm CDM}$ model in~\cite{Eingorn:2015hza}. It is an interesting task to investigate the same for other viable gravity models as well. Being motivated by this, we have investigated the cosmological screening at such large length scales for the phantom braneworld model  described in \ref{s2}, with the expectation that the qualitative differences of this model compared to $\Lambda{\rm CDM}$ should be maximum at the (super-)horizon scales of our universe. We have presented the equations governing the first order scalar and vector perturbations in \ref{s3}. Finally, by ignoring the backreaction effects due to the bulk cosmological constant and the vector perturbation, we have demonstrated analytically and numerically, the behaviour of the two potentials up to the superhorizon length scale in \ref{s4}. 

It seems to be an interesting task to investigate the tensor perturbation for this model in an early universe scenario. We hope to address this issue in future work. 
\section*{Acknowledgements}
\noindent
SB acknowledges M.~Eingorn for mentioning useful references and V.~Sahni for discussions on the phantom braneworld model. SRK wishes to acknowledge the ITCP of the University of Crete for a graduate fellowship. TNT wishes to thank CERN-TH for their hospitality during the late stages of this work.


\end{document}